\documentclass[10pt,aps,prb,twocolumn,superscriptaddress,showpacs]{revtex4-2}

\usepackage[pdftex]{graphicx}
\usepackage{hyperref}
\usepackage{amssymb}
\usepackage{amsmath}
\usepackage{xspace}
\usepackage{color}

\usepackage[dvipsnames]{xcolor}

\newcommand{\TNS}{Ta$_2$NiSe$_5$}
\newcommand{\TNSs}{Ta$_2$NiSe$_5$\ }
\newcommand{\angstrom}{\mbox{\normalfont\AA}}

\begin{document} 

\title{Photoinduced metallization of excitonic insulators}

\author{Satoshi Ejima}
\affiliation{Institut f{\"u}r Physik, Universit{\"a}t Greifswald, 17489 Greifswald, Germany}
\author{Florian Lange}
\affiliation{Institut f{\"u}r Physik, Universit{\"a}t Greifswald, 17489 Greifswald, Germany}
\affiliation{Erlangen National High Performance Computing Center, Friedrich-Alexander-Universit\"at
  Erlangen-N{\"u}rnberg, 91058 Erlangen, Germany}
\author{Holger Fehske}
\affiliation{Institut f{\"u}r Physik, Universit{\"a}t Greifswald, 17489 Greifswald, Germany}
\affiliation{Erlangen National High Performance Computing Center, Friedrich-Alexander-Universit\"at
  Erlangen-N{\"u}rnberg, 91058 Erlangen, Germany}

\date{\today}

\begin{abstract}
Utilizing the time-dependent density-matrix renormalization group technique, we numerically prove 
photoinduced pairing states in the extended Falicov-Kimball model (EFKM) at half filling, 
both with and without internal SU(2) symmetry. In the time-dependent photoemission spectra 
an extra band appears above the Fermi energy after pulse irradiation, indicating an insulator-to-metal transition. 
Even in the absence of the SU(2) structure, the pair correlations are enhanced during the pump, 
and afterwards they decrease over time. This implies the possible metallization of Ta$_2$NiSe$_5$, 
a strong candidate for an excitonic insulator material, for which the EFKM is considered to be the minimal theoretical model. 
Simulating the photoemission with optimized pulse parameters, 
we demonstrate a photoinduced quantum phase transition, in accord with recent findings 
in time- and angle-resolved photoemission spectroscopy experiments on Ta$_2$NiSe$_5$.
\end{abstract}

\maketitle

\section{Introduction}
Applying optical pulses to create new phases of matter or enhance different orders 
is becoming a key tool for the study of complex quantum many-body systems~\cite{Basov17,Ishihara19}.
One of the most fascinating findings in this respect is the photoinduced  
superconductivity in strongly-correlated materials~\cite{Fausti2011,Mitrano2016,Budden2021}.
As to the pairing mechanism, from a theoretical point of view, the \textit{$\eta$-pairing}, 
originally proposed for the fermionic Hubbard model by Yang~\cite{Yang89}, has attracted renewed 
attention~\cite{PhysRevB.94.174503,Kaneko19,PhysRevLett.123.030603,PhysRevB.102.165136}. 
Even though $\eta$-pairing is absent in the ground state of Mott insulators,
it can be enhanced by optical pumping utilizing the symmetry structure of the Hubbard model~\cite{Kaneko19}.
Tuning the pump parameters, a photoinduced transition from the Mott-insulating state to a metallic $\eta$-pairing state
could be numerically confirmed quite recently~\cite{Ejima22}.

Other targets of pump-probe measurements are the excitonic-insulator (EI) candidate materials, 
such as 1$T$-TiSe$_2$~\cite{Rohwer2011,Hellmann2012,Mathias2016} and 
\TNS~\cite{PhysRevLett.119.086401,Okazaki2018,PhysRevB.101.235148,Baldini20}. 
In semimetals or narrow-gapped semiconductors, conduction-band electrons and 
valence-band holes can form bound states, so-called excitons. 
Depending on the bandstructure and  strength of the electron-hole interaction, these excitons might condense in a 
Bardeen-Cooper-Schrieffer (BCS) or Bose-Einstein (BE) type way~\cite{PhysRevB.74.165107,PhysRevB.85.121102}. 
While the EI, triggered by electronic interactions, was proposed theoretically more than 
50 years ago~\cite{Mott1961,Knox1963,KM65,PhysRev.158.462,RevModPhys.40.755}, there has been no unequivocal experimental observation as of yet.  
Since 1$T$-TiSe$_2$ has an indirect gap and its potential EI state is accompanied by 
a band folding analogous to a Peierls insulator, it is difficult to exclude
the action of electron-lattice interactions~\cite{PhysRevB.88.205123}. 
Hence, \TNSs might presently be the only candidate for an (purely electronic) EI and, as a consequence, 
it has been studied extensively in both equilibrium and nonequilibrium situations.
In equilibrium,  a characteristic flattening of the valence-band top was observed by 
angle-resolved photoemission spectroscopy (ARPES), implying a BCS-type 
EI~\cite{PhysRevLett.103.026402,PhysRevB.90.155116,PhysRevB.100.245129}.
Then time- and angle-resolved photoemission spectroscopy (TARPES) enables us to
track a nonequilibrium phase transition induced by a short optical pulse. 
The  increase or decrease of the gap depends on the pump fluence, 
and  it has been intensively discussed as a signature of the build up  of excitonic order~\cite{PhysRevLett.119.086401,Baldini20}.
Moreover, a photoexcited insulator-to-metal transition was observed recently~\cite{Okazaki2018,PhysRevB.101.235148}.

Stimulated by these pump-probe experiments performed for \TNS, several theoretical
studies have been carried out on the extended Falicov-Kimball model (EFKM) and its extensions. 
While they basically rely on GW and mean-field-based 
approximations~\cite{PhysRevB.94.035121,PhysRevLett.119.247601,PhysRevB.98.235127}, 
an unbiased numerical analysis using matrix-product-state (MPS) techniques 
would be also possible due to the (quasi-) one-dimensional (1D) structure of \TNS.  
It recently has been shown that the idea of $\eta$-pairing in the Hubbard model 
can be extended to the minimal theoretical model for \TNS, i.e., the 1D EFKM, 
by employing the time-dependent exact-diagonalization method~\cite{Fujiuchi19}.
Here, the internal SU(2) symmetry of the EFKM is represented by $\Delta$ operators
instead of $\eta$ operators in the Hubbard model, when valence ($f$) and conduction ($c$) electron bands
have the same bandwidth, i.e., $t_f=-t_c$. 
Then, depending on the pump parameters, $\Delta$-pairing correlations can be enhanced after pulse irradiation,
by analogy to pumped Mott insulators. 
This opens a route to realize a photoinduced insulator-to-metal transition in the EFKM when simulating
the nonequilibrium dynamics, bearing in mind TARPES experiments on \TNS. 
Here, an issue is that  \TNSs can not be described by the EFKM with internal SU(2) structure since  $t_f\neq -t_c$
(note that with decreasing  $|t_f/t_c|$, pair correlations are strongly suppressed after pulse irradiation~\cite{Fujiuchi19}). 
Hence, an approximation-free large-scale numerical analysis 
is desirable to prove or disprove whether the photoexcited phase transition really occurs  for the parameter set describing  \TNS.

To this end, we investigate the pulse-excited states of the 1D EFKM exploiting the rather unbiased time-dependent DMRG (t-DMRG)~\cite{White92,PhysRevLett.93.076401,Daley2004,Sch11} 
and time-evolving block decimation (TEBD)~\cite{TEBD} techniques. To eliminate finite-size effects the
pair correlations will be simulated directly in the thermodynamic limit ($L\to\infty$) 
by means of the infinite TEBD (iTEBD)~\cite{iTEBD,Reortho} technique based on 
an infinite MPS (iMPS) representation. 
The nonequilibrium dynamics of the driven system is computed by the t-DMRG method with so-called
infinite boundary conditions (IBCs)~\cite{Zauner2015}. 

The paper is organized as follows. In Sec.~\ref{model-method}, we first introduce the EFKM and then briefly describe our numerical method.
In Sec.~\ref{results}, we determine the optimal pulse parameter in the SU(2) symmetric case ($t_f=-t_c$)
and we demonstrate the photoinduced phase transition. Then, we examine 
the possible transient pairing state for $t_f\neq -t_c$, in view of \TNS. 
Our main conclusions are presented in Sec.~\ref{summary}.

\section{Model and method}
\label{model-method}

\subsection{Extended Falicov-Kimball model}
Because of the quasi-1D lattice structure (note that significant electron hopping 
only takes place along the parallel Ni and Ta chains), the EFKM~\cite{BGBL04,PhysRevB.81.115122,2D-EFKM} 
in one dimension~\cite{1D-EFKM,Ejima21} is believed to be the minimal theoretical model 
for \TNS~\cite{PhysRevB.84.245106,PhysRevB.90.155116}.
The Hamiltonian of the 1D EFKM  reads
\begin{align}
  \hat{H}= &
     -\sum_{\alpha=c,f} t_\alpha\sum_{j}
     \left(\hat{\alpha}_{j}^\dagger \hat{\alpha}_{j+1}^{\phantom{\dagger}}
      +\text{H.c.}\right)
     +U\sum_j\hat{n}_j^{c}\hat{n}_j^{f}
     \nonumber\\
    &+\frac{D}{2}\sum_j\left(\hat{n}_j^{c}-\hat{n}_j^{f}\right)
    -\mu\sum_{\alpha,j}\hat{n}_j^{\alpha}
    \,,
  \label{EFKM}
\end{align}
where $\hat{\alpha}_j^{\dagger}$ ($\hat{\alpha}_j^{\phantom{\dagger}}$) is the creation (annihilation) 
operator of a spinless fermion in the $\alpha=\{c,f\}$ orbital at Wannier site $j$, 
$\hat{n}_j^{\alpha}=\hat{\alpha}_j^{\dagger}\hat{\alpha}_j^{\phantom{\dagger}}$, 
and $U$ is the local Coulomb repulsion between $c$ and $f$ electrons staying at the same lattice site.
$D$ parametrizes the level splitting of $c$ and $f$ orbitals, and $\mu$ is the chemical potential.
At fixed $|t_f/t_c|$, the phase diagram of the EFKM in the $D$-$U$ plane~\cite{BGBL04,1D-EFKM} exhibits 
three insulating phases: the staggered orbital ordered phase corresponding to an antiferromagnetic state of the Hubbard model 
in the strong-coupling limit, the band insulator phase with the mean densities 
$\langle\hat{n}_j^c\rangle=0$ and $\langle\hat{n}_j^f\rangle=1$,
and the EI phase in between with $0<\langle\hat{n}_j^c\rangle<1/2$ and $1/2<\langle\hat{n}_j^f\rangle<1$,
where the excitonic correlations decay with power law. Dealing with a 1D system, the EI state should be more properly characterized by  the condensation amplitude $F(k)=\langle \psi_1|\hat{c}_k^\dagger \hat{f}_k^{\phantom{\dagger}}|\psi_0\rangle$
in momentum space~\cite{1D-EFKM,Ejima21}. 
Here, $|\psi_0\rangle$ is the ground state for a finite system with $L$ sites and $N_f$ ($N_c$)
$f$-electrons ($c$-electrons),  and $|\psi_1\rangle$ is the excited state with ($N_f-1$) $f$-electrons and 
($N_c+1$) $c$-electrons.  Due to "Fermi surface" effects $F(k)$ exhibits a sharp peak 
at the Fermi momentum $k_{\rm F}=\pi\langle\hat{n}_j^c\rangle$ in the BCS-type EI regime 
where electron-hole pairs are only weakly bound. On the other hand, $F(k)$ shows a maximum
at $k=0$ for tightly bound excitons in the BE condensate (BEC)-type EI regime. In this way, one can successfully define 
a BCS-BEC crossover region, where $F(k)$ has a maximum for $0<k<k_{\rm F}$, 
as demonstrated in Ref.~\cite{Ejima21}.

In the following, in order to explore the effects of photoexcitation, we concentrate on the EI state.

Let us first define the so-called $\Delta$-pairing operators~\cite{Fujiuchi19}:
\begin{align}
 \hat{\Delta}^+&=\sum_j \hat{c}_j^\dagger\hat{f}_j^\dagger\equiv\sum_j\hat{\Delta}_j^+\,,
 \ \ \
 \hat{\Delta}^-=\left(\hat{\Delta}^+\right)^\dagger\,,
 \\
\hat{\Delta}^z&=\frac{1}{2}\sum_j(\hat{n}_j^f+\hat{n}_j^c-1)\equiv\sum_j\hat{\Delta}_j^z\,,
\end{align}
which satisfy SU(2) commutation relations, i.e.,
\begin{align}
 [\hat{\Delta}^+,\hat{\Delta}^-]=2\hat{\Delta}^z\,,
 \ \
 [\hat{\Delta}^z,\hat{\Delta}^\pm]=\pm\hat{\Delta}^\pm\,.
\end{align}
The Hamiltonian of the EFKM~\eqref{EFKM} commutes with the total $\Delta$-pairing operator
\begin{align}
 \hat{\Delta}^2=\frac{1}{2}(\hat{\Delta}^+\hat{\Delta}^-+\hat{\Delta}^-\hat{\Delta}^+)
  +(\hat{\Delta}^z)^2
\end{align}
if $t_f=-t_c$, so that $\langle\hat{\Delta}^2\rangle$ is a conserved quantity
in the absence of perturbations. Eigenstates with a finite value of $\langle\hat{\Delta}^2\rangle$
have long-ranged pairing correlations $\langle\hat{\Delta}_j^+\hat{\Delta}_\ell^-\rangle$
analogous to the $\eta$-pairing correlations in the Hubbard model.
While $\Delta$-pairing states do not appear in the various ground states of the EFKM, they can be induced by driving the system out of equilibrium by pulse irradiation.

In this study, we demonstrate the existence of such photoinduced $\Delta$-pairing states monitoring the time-evolution. We introduce an external time-dependent electric field $A(t)$ via a Peierls phase
in the first term of Eq.~\eqref{EFKM},
$t_\alpha\hat{\alpha}_{j}^\dagger\hat{\alpha}_{j+1}^{\phantom{\dagger}}
\to t_\alpha e^{\mathrm{i}A(t)}\hat{\alpha}_{j}^\dagger\hat{\alpha}_{j+1}^{\phantom{\dagger}}$,
where
\begin{align}
  A(t)=A_0 e^{-(t-t_0)^2/2\sigma_{\rm p}^2}\cos[\omega_{\rm p}(t-t_0)]\,,
\end{align}
which describes a pump pulse with amplitude $A_0$, frequency $\omega_{\rm p}$ and
width $\sigma_{\rm p}$, centered at time $t_0$ ($>0$).  As a result, the Hamiltonian becomes
time dependent, $\hat{H}\to\hat{H}(t)$, and the system is driven out of equilibrium, evolving in time out of the ground state as $|\psi(0)\rangle\to|\psi(t)\rangle$.

Hereafter we use $t_c$ ($t_c^{-1}$) as the unit of energy (time) and consider
the half-filling case, $L=N(=\sum_\alpha N_\alpha)$. Note that the number of electrons
$N_\alpha$ of each orbital $\alpha$ is conserved during pulse irradiation.

\subsection{Numerical method}
\label{method}
The $\Delta$-pairing state can be detected by analyzing the time evolution of
the pair-correlation function
\begin{align}
 P(r,t)=\frac{1}{L}\sum_j\langle\psi(t)|
  (\hat{\Delta}_{j+r}^+\hat{\Delta}_{j}^-+{\rm H.c.})
  |\psi(t)\rangle
\end{align}
and its Fourier transform $\tilde{P}(q,t)=\sum_r e^{\mathrm{i}qr}P(r,t)$.

Because of the strong electron-hole correlation in the ground state of the EFKM,
it is also of specific importance to estimate the time evolution of the excitonic
correlation function defined with the exciton creation operators 
$\hat{b}_j^{\dagger}=\hat{c}_j^\dagger \hat{f}_j^{\phantom{\dagger}}$ as
\begin{align}
 N_{\rm ex}(r,t)=\frac{1}{L}\sum_j\langle \psi(t)|
 (\hat{b}_{j+r}^\dagger\hat{b}_{j}^{\phantom{\dagger}}+{\rm H.c.})
 |\psi(t)\rangle\,,
\end{align}
and its Fourier transform $\tilde{N}_{\rm ex}(q,t)=\sum_r e^{\mathrm{i}qr}N_{\rm ex}(r,t)$.
As demonstrated in Ref.~\cite{Fujiuchi19} for small clusters in the EFKM with $t_f=-t_c$,
$\tilde{P}(q=0,t)$ is enhanced after pulse irradiation, indicating the formation of
a $\Delta$-pairing state. At the same time, the excitonic correlations $\tilde{N}_{\rm ex}(q=0,t)$
will be strongly suppressed. The question is whether this is also the case for $t_f\neq -t_c$,
i.e., in the absence of SU(2) symmetry.

Before we answer this question, we briefly outline how we proceed simulating the nonequilibrium dynamics in the EFKM.
The photoemission spectrum for orbital $\alpha$ is
\begin{align}
  A_\alpha^-(k,\omega;t) = \sum_{r}e^{-\mathrm{i}kr} \int_{-\infty}^{\infty}\int_{-\infty}^{\infty} d\tau_1 d\tau_2 f(\tau_1,\tau_2;\omega)
  \nonumber \\
  \times C_\alpha(r,\tau_1,\tau_2; t) 
  \label{noneq-dyn}
\end{align}
with the two-point correlator at times $\tau_1$ and $\tau_2$ (defined relative to time $t$)
\begin{align}
  C_\alpha(r,\tau_1,\tau_2; t)=
  \langle\phi(t)|\hat{\alpha}_{j+r}^\dagger(\tau_1;t)
                \hat{\alpha}_{j}^{\phantom{\dagger}}(\tau_2;t)|\phi(t)\rangle\,.
\end{align}
The prefactor in Eq.~\eqref{noneq-dyn},
$f(\tau_1,\tau_2;\omega)=e^{{\mathrm i}\omega(\tau_1-\tau_2)}g(\tau_1)g(\tau_2)$ with
$g(\tau)=\exp[-\tau^2/2\sigma_{\rm pr}^2]/\sqrt{2\pi}\sigma_{\rm pr}$, describes the shape of a probe pulse.
We calculate $C(r,\tau_1,\tau_2; t)$ numerically by simulating states
$\hat{\alpha}_j(\tau;t)|\phi(t)\rangle=\hat{U}^\dagger(t+\tau,t) \hat{\alpha}_j \hat{U}(t+\tau,t)|\phi(t)\rangle\equiv |\bar{\phi}(\tau;t)\rangle$. 
Here,
$\hat{U}(t+\tau,t)\equiv {\cal T}\exp\left[-\mathrm{i}\int_t^{t+\tau} dt^\prime \hat{H}(t^\prime)\right]$
is the unitary time-evolution operator of the system with the (reverse) time-ordering operator ${\cal T}$
for $\tau>0$ ($\tau<0$).
The wave functions $|\bar{\phi}(\tau;t)\rangle$ in two time domains, simulated 
with IBCs~\cite{Zauner2015}, provide us with the two-point correlators $C(r,\tau_1,\tau_2; t)$. 
Note that the computational efforts for the simulations of $C(r,\tau_1,\tau_2; t)$ can be reduced 
by utilizing the iMPS unit cells, as explained in Refs.~\cite{Spin1FinT,Ejima21} (see also Ref.~\cite{Ejima22} for further details).

In the TEBD (iTEBD) calculations, we employ a second-order Suzuki-Trotter decomposition with time step
$0.1t_c^{-1}$ ($0.01t_c^{-1}$). The maximum used bond dimension is 1600,
which ensures that truncation error is smaller than $10^{-5}$. 
As the time cutoff in the simulation of the time-dependent correlation functions we choose $T=5t_c^{-1}$,
so that the integration in Eq.~\eqref{noneq-dyn} is done over
$-T\leq\tau_1,\tau_2\leq T$.
To find a compromise between time and frequency resolutions, we choose a probe width 
$\sigma_{\rm pr}\cdot t_c=2.0$.

\section{Numerical results}
\label{results}

\subsection{SU(2)-symmetric EFKM}
\begin{figure}[tb]
  \includegraphics[width=0.99\linewidth]{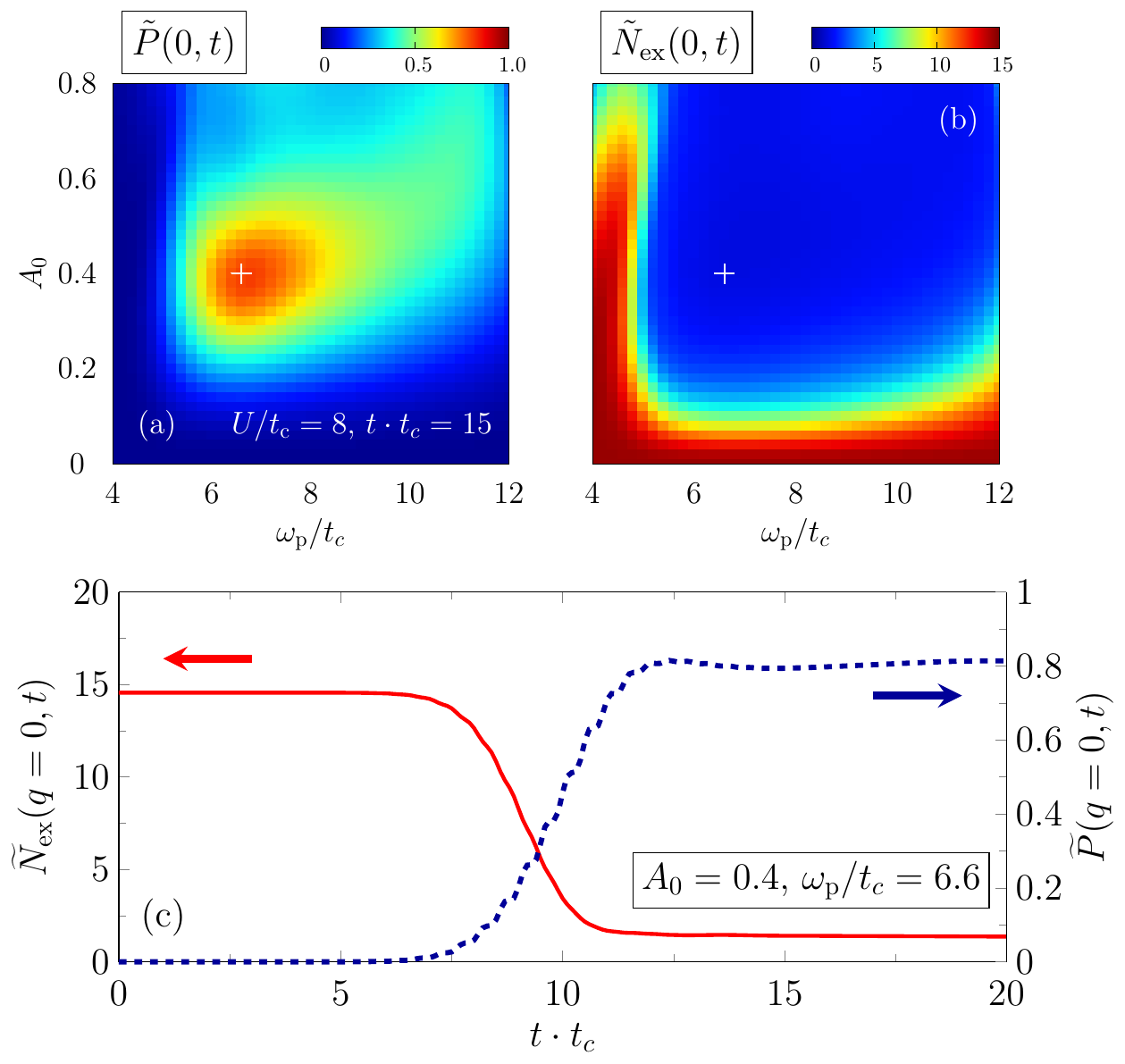}
  \caption{
   Contour plots of $\tilde{P}(q=0,t)$ (a) and $\tilde{N}_{\rm ex}(q=0,t)$ (b) at $t\cdot t_c=15$ 
   in the $\omega_{\rm p}$-$A_0$ plane for an infinite half-filled EFKM chain 
   with $t_f=-t_c$, $U/t_c=8$,  $\langle \hat{n}_c\rangle=0.25$, and $D/t_c=0.75$. 
   (c): Time evolution of $\tilde{P}(0,t)$ and $\tilde{N}_{\rm ex}(0,t)$ near the peak
   position of panel (a), $A_0=0.4$ and $\omega_{\rm p}/t_c=6.6$ marked by the `+' symbol
   in panels (a) and (b). 
   The pump pulse has width $\sigma_{\rm p}\cdot t_c=2$ and is centered at time $t_0\cdot t_c=10$. 
   }
   \label{contour-plots1}
\end{figure}

Let us first investigate the SU(2) symmetric case where $t_f=-t_c$ in the EFKM. 
Using the time-dependent exact-diagonalization technique, Fujiuchi et al. showed the enhancement of 
the $\Delta$-pair correlations and the suppression of the excitonic correlations 
after pulse irradiation in this model~\cite{Fujiuchi19}.
Because of the small size of the considered clusters ($L=16$), however, the $A_0$ and $\omega_{\rm p}$ dependence of the $\Delta$-pair correlations was found to exhibit an artificial stripe structure. 

\begin{figure*}[tb]
  \includegraphics[width=0.9\linewidth]{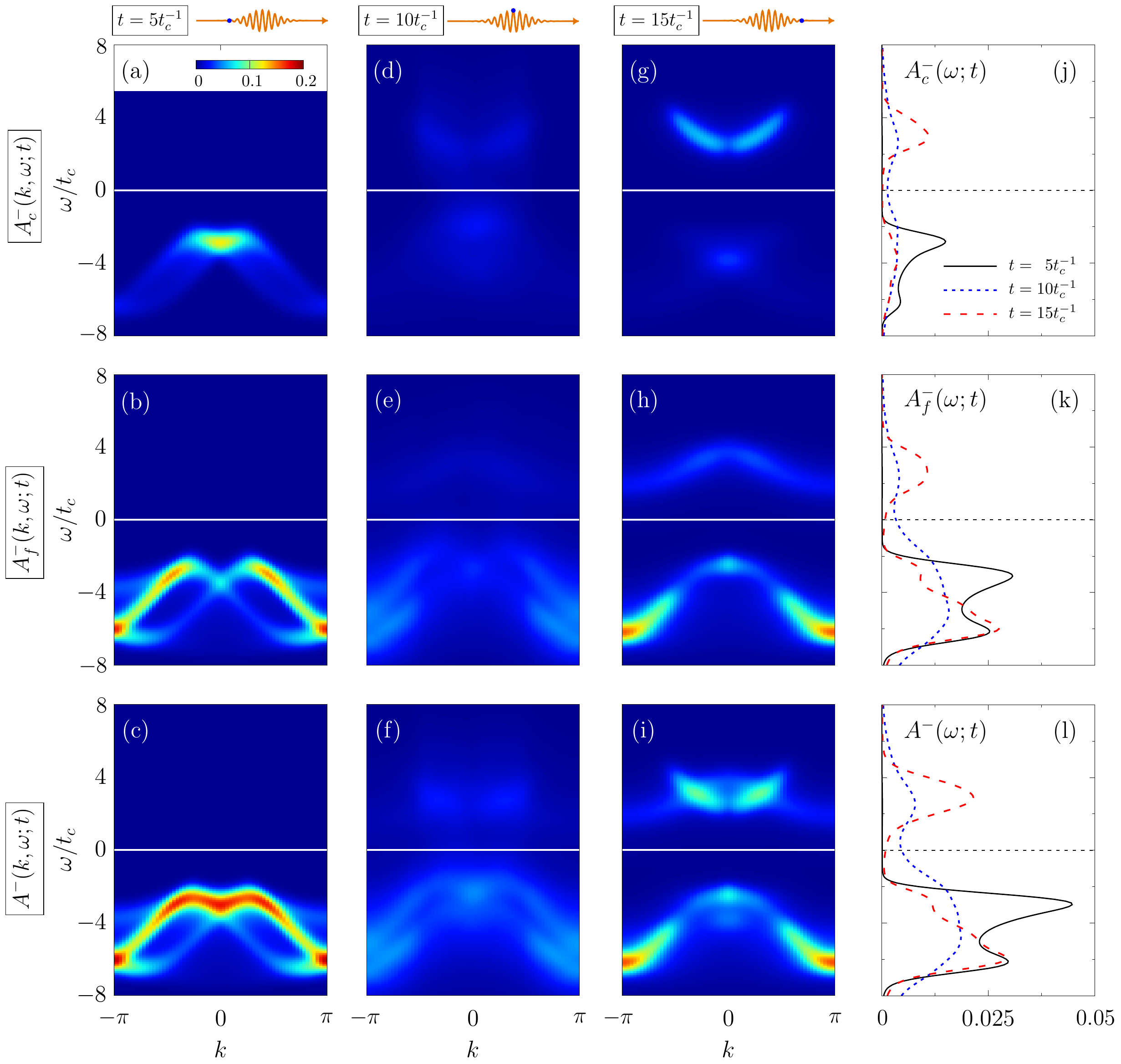}
  \caption{(a)-(i): Snapshots of the photoemission spectra $A_{\alpha=c,f}^-(k,\omega;t)$ and 
  $A^-(k,\omega;t)=\sum_\alpha A_{\alpha}^-(k,\omega;t)$ for the $\Delta$-pairing 
  dominant state. Here, $A_0=0.4$ and $\omega_{\rm p}/t_c=6.6$ denoted as `+' in Fig.~\ref{contour-plots1}.  
  Results are displayed during the pump at $t\cdot t_c=5$ [(a)-(c)],  10 [(d)-(f)], and 15 [(g)-(i)]. 
  (j)-(l): Transient integrated density of states $A_{(\alpha)}^-(\omega;t)$ from Eq.~\eqref{dos} 
  obtained from the results in (a)-(i).
  All data obtained by the (i)TEBD technique with IBCs,
  with the pump pulse parametrized as in Fig.~\ref{contour-plots1}.
  The Fermi energy $E_{\rm F}$ is set to $\omega=0$. 
  }
  \label{TARPES1}
\end{figure*}

By calculating $\tilde{P}(q=0,t)$ directly in the thermodynamic limit ($L\to\infty$) using iTEBD this finite-size effect can be eliminated. This is demonstrated by Fig.~\ref{contour-plots1}(a). The obtained single-peak structure is similar to that observed for the $\eta$-pairing states in the half-filled Hubbard model~\cite{Ejima20,Ejima22}. 
In the parameter region where the electron-electron pair correlations $\tilde{P}(0,t)$ are enhanced, the excitonic correlations 
$\tilde{N}_{\rm ex}(q=0,t)$ [Fig.~\ref{contour-plots1}(b)] are strongly suppressed. 
As shown in Fig.~\ref{contour-plots1}(c), this behavior becomes more evident when focusing on the time evolution of both correlation functions for the parameter set which leads to the strongest increase of $\tilde{P}(0,t)$, i.e., 
$A_0=0.4$ and $\omega_{\rm p}/t_c=6.6$ [denoted by `+' in Fig.~\ref{contour-plots1}(a)].
After pulse irradiation ($t\gtrsim t_0$), $\tilde{P}(0,t)$ grows rapidly in time and saturates to a value of about $0.8$, while $\tilde{N}_{\rm ex}(0,t)$ is strongly suppressed for $t\gtrsim t_0$.

The enhancement process of pair correlations can be described as follows~\cite{Fujiuchi19}:
The initial state before pulse irradiation is the ground state
of the EFKM $\hat{H}$ with the eigenstate $|\Delta=0,\Delta^z=0\rangle$,
which is consistent with the numerical result $\tilde{P}(0,t=0)\approx 0$ in Fig.~\ref{contour-plots1}(c). 
[When $t_f=-t_c$, any eigenstate of $\hat{H}$ can be
represented by an eigenstate of $\hat{\Delta}^2$ and $\hat{\Delta}^z$ with eigenvalues $\Delta(\Delta+1)$ and $\Delta^z$, respectively.]
Turning on the pump pulse, the commutation relation between the Hamiltonian and $\Delta$ operators is broken,
\begin{align}
 [\hat{H}(t), \hat{\Delta}^+]=[\hat{H},\Delta^+]\cos[A(t)] +\sum_k F(k,t)\hat{c}_{-k}^\dagger\hat{f}_k^\dagger\,,
\end{align}
where $F(k,t)=4t_f\sin[A(t)]\sin k$, provided that $t_f=-t_c$. 
This brings the initial state to one with a finite expectation value 
$\langle \hat{\Delta}^2 \rangle$. For $t\gg t_0$, the commutation relation is recovered, i.e., 
$[\hat{H}(t), \hat{\Delta}^+]\to[\hat{H}, \hat{\Delta}^+]$, 
since $A(t)\to0$ for large $t$. 
However, $|\psi(t)\rangle$ now includes components of $|\Delta>0,\Delta^z=0\rangle$, leading to the enhancement of $\tilde{P}(0,t)$.
Calculating all eigenstates as well as $\tilde{P}(0,t)$ for a small cluster by full exact diagonalization and taking the selection rule of $\Delta$ pairs into account, Fujiuchi et al. could indeed show that the photoinduced state $|\psi\rangle$ is related to the $\Delta$-pairing state~\cite{Fujiuchi19}.

We now analyze the nonequilibrium photoemission spectra for the optimal parameter set 
denoted as `+' in Fig.~\ref{contour-plots1}, in order to discuss the  photoinduced insulator-to-metal transition in  the EFKM in close analogy to the Mott-insulator--to--$\eta$-pairing-state transition in the half-filled Hubbard model~\cite{Ejima22}.

Figure~\ref{TARPES1} displays our results  for the 1D half-filled EFKM with SU(2) symmetry. 
Before pump irradiation the state is an EI with a significant single-particle gap, as confirmed by Figs.~\ref{TARPES1}(a)-(c) for $t\cdot t_c=5$. 
In the middle of the pump ($t\cdot t_c=10$), the spectral weights is shifted 
both below and above the Fermi energy $E_{\rm F}$ and the band dispersions become smeared [Figs.~\ref{TARPES1}(d)-(f)].
Figures~\ref{TARPES1}(g)-(i) indicate that the additional (dispersive) signals persist above $E_{\rm F}$  after pulse irradiation ($t\cdot t_c=15$).

Extracting the integrated density of states from the photoemission spectra,
\begin{align}
 A_{(\alpha)}^-(\omega;t)=\frac{1}{L}\sum_k A_{(\alpha)}^-(k,\omega;t)\,,
 \label{dos}
\end{align}
makes it easier to see how the spectral weight is shifted from $\omega<E_{\rm F}$ to 
$\omega>E_{\rm F}$ by the photoinduced $\Delta$-pairing. Obviously, the spectral weight 
for $\omega>E_{\rm F}$ in Figs.~\ref{TARPES1}(j)-(l) increases over time for both $c$ and $f$ orbitals, indicating the photoinduced metallization of the EI in the course of $\Delta$-pairing. 

Let us emphasize that in the $\Delta$-pairing nondominant regime the photoinduced transition will not occur, i.e., the photoemission spectra are barely changed even after pulse irradiation, just like in pumped Mott insulators~\cite{Ejima22}.

\subsection{EFKM for \TNS}
\begin{figure}[tb]
  \includegraphics[width=0.99\linewidth]{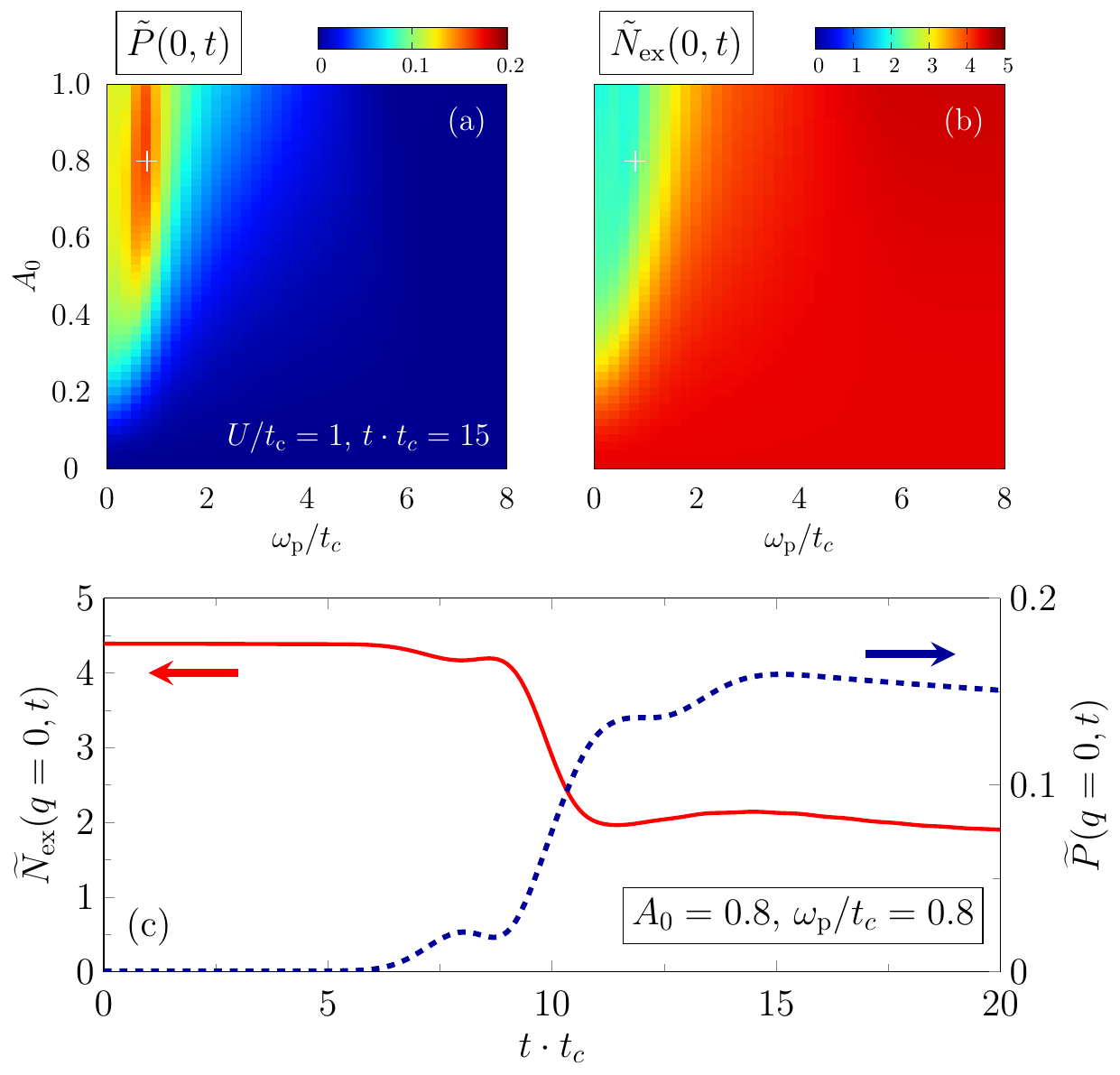}
  \caption{
   Contour plots of $\tilde{P}(0,t)$ (a) and $\tilde{N}_{\rm ex}(0,t)$ (b) 
   in the $\omega_{\rm p}$-$A_0$ plane for an infinite EFKM chain after pulse irradiation ($t\cdot t_c=15$).
   The EFKM model parameters are  $t_f/t_c=-0.5$, $U/t_c=1$, $D/t_c=0.03$, and $\langle \hat{n}_j^c \rangle=0.1$.
   (c): Time evolution of $\tilde{P}(0,t)$ and $\tilde{N}_{\rm ex}(0,t)$
   near the peak position of panel (a), i.e., $A_0=0.8$ and $\omega_{\rm p}/t_c=0.8$
   marked by the `+' symbol in panels (a) and (b).
   The pump pulse has width $\sigma_{\rm p}\cdot t_c=2$ and is centered at time $t_0\cdot t_c=10$.
   }
   \label{contour-plots05}
\end{figure}

Finally we explore the electron-electron pair correlations in the photoexcited state of our target material \TNS. 
At the first glance, the above discussion does not seem applicable to  \TNSs where the SU(2) symmetry is broken. 
The optimal parameter set for an EFKM-based description of \TNSs can be determined from ARPES data. 
Considering a three-chain electron-phonon-coupled system and carrying out a  bandstructure calculation
supplemented by a mean-field analysis, the (EFKM) model parameters are estimated as $t_c=0.8$, $t_f=-0.4$, 
$U\simeq 0.55$ and $D=0.2$ in units of eV~\cite{ParaEstimation-Kaneko}. 
So clearly, $t_f\neq -t_c$ and the commutation relations with respect to the $\Delta$-pairing operators are broken, 
i.e., $[\hat{H},\hat{\Delta}^\pm]\neq \pm U\hat{\Delta}^\pm$ even without any auxiliary perturbation. 
$\tilde{P}(0,t)$ is therefore not a conserved quantity at any time and, in contrast to the SU(2)-symmetric case, decreases after the pulse irradiation. The suppression of $\tilde{P}(0,t)$ becomes stronger with decreasing $|t_f/t_c|$. 
Thus, the first question is whether $\tilde{P}(0,t)$ is sufficiently enhanced at least during the transient
period to realize the photoinduced insulator-to-metal transition 
observed in the  \TNS-TARPES experiments~\cite{Okazaki2018,PhysRevB.101.235148}.

Moreover, it is worth pointing out that the band flattening detected in the ARPES experiments for \TNSs occurs only in a relatively narrow region of momentum space, namely for the momentum $|k|\lesssim 0.1\angstrom^{-1}$. 
This means $|k|/a\lesssim 0.35$, taking into account that the lattice constant of the chain direction is $a=3.51232\; \angstrom$~\cite{Jain2013}. 
Accordingly, for $t_f/t_c=-0.5$ and $U/t_c\lesssim 1$, the flat band can appear for 
$|k|\lesssim k_{\rm F}$, since the system is in the EI-BCS regime as demonstrated in Ref.~\cite{Ejima21}.
Therefore the magnitude of the level splitting $D$ takes values such that the system is also very close to the excitonic-to-bond insulator transition line,
i.e., for  $\langle\hat{n}_j^c\rangle=0.1$ we have $k_{\rm F}=0.1\pi(<0.35)$~\cite{Ejima21}. 
Hence, the second issue is that a high momentum resolution in the calculation of nonequilibrium spectral functions is mandatory, which cannot be achieved by time-dependent exact diagonalizations [being very limited with respect to the tractable system sizes ($L\lesssim16$ at present)]. 
The t-DMRG technique with IBCs, however, enables the desired high-resolution analysis of the nonequilibrium dynamics (cf. the results for the half-filled Hubbard chain~\cite{Ejima22}). 

\begin{figure*}[tb]
  \includegraphics[width=0.9\linewidth]{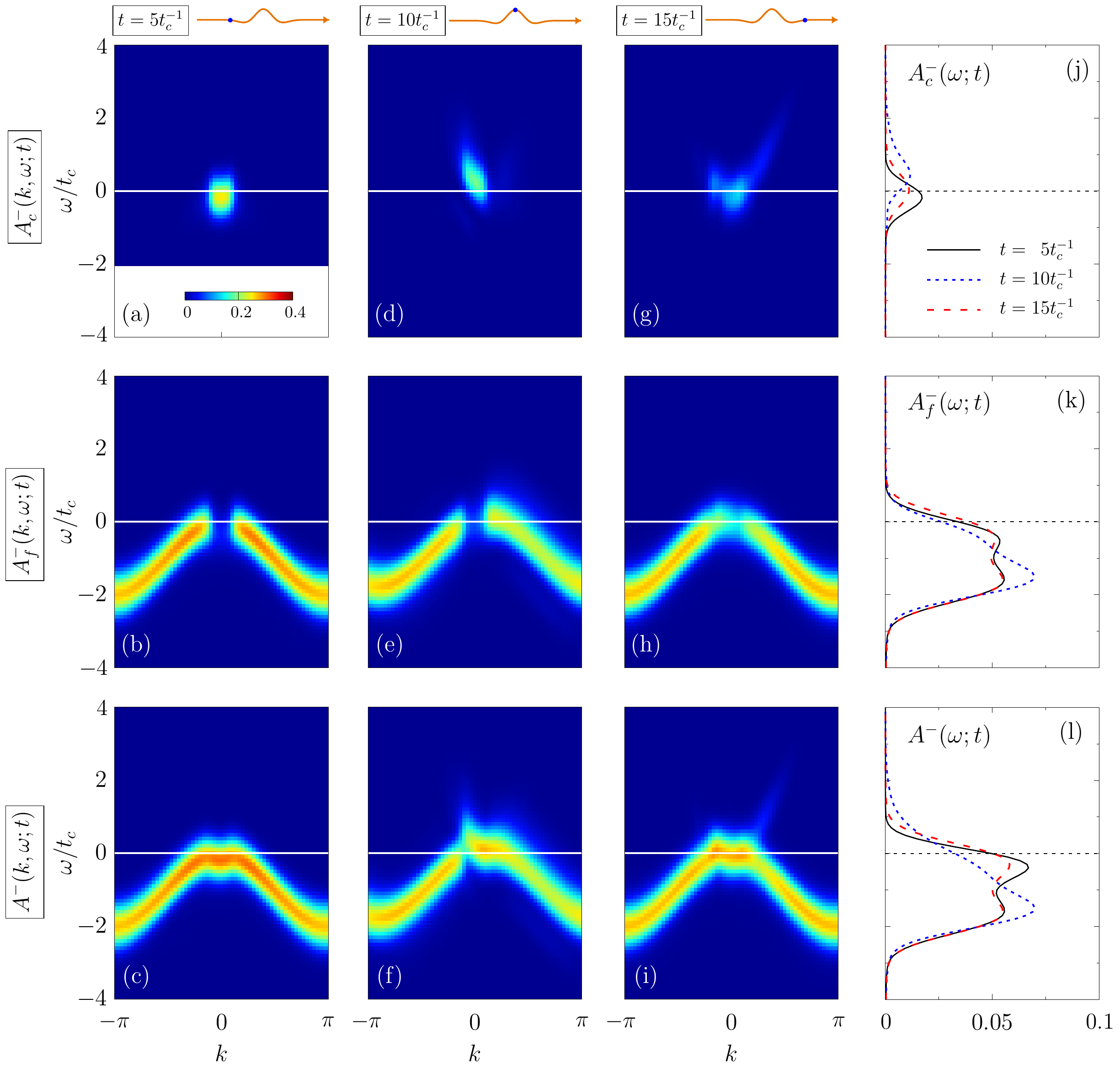}
  \caption{(a)-(i): Snapshots of the photoemission spectra $A_{\alpha=c,f}^-(k,\omega;t)$ and
  $A^-(k,\omega;t)=\sum_\alpha A_{\alpha}^-(k,\omega;t)$ of the 1D EFKM with $U/t_c=1$ and $\langle \hat{n}_c\rangle=0.1$. 
  Results are given, for the $\Delta$-pairing dominant state, during the pump at $t\cdot t_c=5$ [(a)-(c)],  10 [(d)-(f)], and 15 [(g)-(i)], where 
  $A_0=0.8$ and $\omega_{\rm p}/t_c=0.8$ (marked by the `+' in Fig.~\ref{contour-plots05}). 
  (j)-(l): The transient integrated density of states $A_{(\alpha)}^-(\omega;t)$ obtained from panels (a)-(i).
 All  data  obtained by the (i)TEBD technique with IBCs.
  }
  \label{TARPES05}
\end{figure*}

Let us first examine whether the pair correlations $\tilde{P}(0,t)$ are enhanced 
for the estimated  \TNSs parameter set after pulse irradiation by tuning the pump characteristics. 
Figure~\ref{contour-plots05}(a) displays the contour plot of  $\tilde{P}(0,t)$ in the EFKM
with $t_f/t_c=-0.5$ after the light pulse ($t\cdot t_c=15$). Obviously, an enhancement of 
$\tilde{P}(0,t)$ appears around $\omega_{\rm p}\approx U$, for large enough amplitudes $A_0\gtrsim0.6$.
At the same time, again, the excitonic correlations $\tilde{N}_{\rm ex}(0,t)$ are strongly suppressed in this
regime, see Fig.~\ref{contour-plots05}(b). 
The time evolution of both correlation functions is shown in Fig.~\ref{contour-plots05}(c) 
for the optimal parameter set of $\tilde{P}(0,t)$
[$A_0=0.8$ and $\omega_{\rm p}/t_c=0.8$ denoted as `+' in Fig.~\ref{contour-plots05}(a)].
After pulse irradiation $\tilde{P}(0,t)$ decreases gradually over time because of
the now missing internal SU(2) symmetry.

In the following, we show that for such optimal pumping a transient $\Delta$-pairing state will be induced. 
Figure~\ref{TARPES05} exhibits nonequilibrium single-particle excitation spectra $A_{(\alpha)}^-(k,\omega;t)$
and the integrated density of states $A_{(\alpha)}^-(\omega;t)$ for the optimal pumped EFKM, using  the \TNSs parameters.
Before the main pulse arrives ($t\cdot t_c=5$), the behavior of the spectral functions is clearly similar to that of 
the equilibrium ones at $T=0$ (cf.~\cite{Ejima21}). 
Namely, $A_c^-(k,\omega;t)$ and $A_f^-(k,\omega;t)$ follow the unrenormalized $c$- and $f$-band dispersions
respectively, reflecting the weakly-coupled electron-hole pairs in a BCS-type EI. 
Only the bottom of the $c$ band appears for $|k|\lesssim  k_{\rm F}(\simeq\pi\langle\hat{n}_j^c\rangle)$
[Fig.~\ref{TARPES05}(a)], while the top of the $f$ band is missing as in Fig.~\ref{TARPES05}(b). 
This gives rise to an ``M''-shaped bandstructure with two peaks close to $\pm k_{\rm F}$ [Fig.~\ref{TARPES05}(c)], 
just as was found in the \TNS-ARPES experiments in the vicinity of the $\Gamma$ point. 
Note that an exponentially-small gap at $k=\pm k_{\rm F}$ ($\Delta_{\rm c}/t_c\approx0.125$) is difficult to detect in panels (a)-(c)
because of the probe width $\sigma_{\rm pr}\cdot t_c=2.0$ as a compromise between time and 
frequency. 
During pump irradiation ($t\cdot t_c=10$) the small but finite spectral weight shifts to the region 
$\omega>E_{\rm F}$ [Figs.~\ref{TARPES05}(d)-(f)] due to the enhanced pairing correlations 
$\tilde{P}(0,t)$. This is another indication that the photoinduced transient metallic state found in the EFKM can be related to 
the \TNS-TARPES experiments.   Focusing on the momentum dependence, we observe that the dispersions 
are asymmetric with respect to $k=0$, where the spectral weight in both $c$ and $f$ orbitals is transferred to 
the negative momentum region.
This can be understood by considering the noninteracting bandstructure within an external field $A(t)$. 
Due to $A(t)$ the bare cosine-band $\epsilon(k)\propto\cos(k)$ becomes 
$\epsilon^\prime(k)\propto\cos(k+A(t))$. $A(t)$ takes its maximum value when $t=t_0$ and the cosine-band 
is shifted to the region with the most negative momentum. 
After pulse irradiation ($t\cdot t_c=15$), the spectral weight shifts back to the positive momentum region, and $\epsilon^\prime(k)\to\epsilon(k)$ for $t\gg t_0$
[see Figs.~\ref{TARPES05} (g)-(i)]. Note that $A_{(\alpha)}^-(\omega;t)$ displayed in panels (j)-(l) shows that
the total spectral weight above $E_{\rm F}$ is decreasing with time, simply because the pair correlation $\tilde{P}(0,t)$
is reduced [see Fig.~\ref{contour-plots05}(c)]. 
It should be emphasized that such an asymmetric behavior is only observed when applying the external field 
in the weak (electron-hole) interaction regime, not for the strongly coupled system, see, e.g., Fig.~\ref{TARPES1}. 
When simulating the photoemission spectra by t-DMRG at finite temperatures without any pulse 
the extracted dispersions are always symmetric with respect to the momentum $k=0$, even in the weak-coupling regime~\cite{Ejima21}. 
Hence, this asymmetry might be used to isolate the photoexcitation effects in time-dependent photoemission from those of the effective temperature increase after pulse irradiation. 

We finally note that the ratio ($\simeq 0.103$) between the observed band gap (0.16 eV)~\cite{Lu2017,PhysRevB.95.195144} and the energy of the laser light (1.55 eV) used in TARPES experiments~\cite{PhysRevLett.119.086401,Okazaki2018} is in reasonable agreement with that obtained from our numerics, $\Delta_{\rm c}/\omega_{\rm p}\simeq 0.156$.

\section{Conclusions}
\label{summary}
To sum up, we have demonstrated the existence of a photoinduced  metallization transition from the excitonic insulator to the $\Delta$-pairing state
in the framework of the half-filled extended Falicov-Kimball model (EFKM),  with and without internal SU(2) symmetry, at zero temperature.  This result, obtained using an unbiased numerical approach, holds for the infinite one-dimensional driven system. 
In the case in which the internal SU(2) structure ($t_f=-t_c$) exists, the finite-size effects inherent to previous small cluster studies could be successfully eliminated
by simulating the relevant two-point correlators directly in the thermodynamic limit.  Importantly, the optimal optical pulse that leads to an enhancement of 
the pair correlations in the course of the insulator-to-metal transition after pulse irradiation has been determined, 
depending on the bandstructure and interaction parameters of the EFKM.

With a view to experiments on \TNS, for which the EFKM with $t_f\neq-t_c$ is considered to be the minimal theoretical model, 
the photoinduced quantum phase transition for optimal pulse parameters is found in agreement with recent TARPES experiments,  even 
though the pair correlations $\tilde{P}(0,t)$ decay over time in the EFKM according to the broken internal SU(2) symmetry if  
$t_f\neq-t_c$.  Most notably, after pulse irradiation the spectral functions exhibit not only the transient 
metallic behavior but also a characteristic asymmetry, which enables us to distinguish the effects of photoexcitation related to 
bandstructure and interaction strength from those of the effective temperature increase due to pulse irradiation. 
Hence this asymmetry,  being absent in the strong-coupling regime, yields further evidence that \TNSs is represented by 
the weak-coupling EFKM, implying (if present) an excitonic insulator phase of BCS-type .
We note that previous \TNSs experiments have still not succeeded in observing this asymmetry directly, 
maybe because of suboptimal pulses.
Therefore it would be very interesting to carry out TARPES experiments with the proposed optimal tuned pulses for enhanced paring states.

\section*{Acknowledgments}
We acknowledge enlightening discussions with T.~Kaneko and Y.~Ohta.  
The (i)TEBD simulations were performed using the ITensor library~\cite{ITensor}. 
S.E. was supported by Deutsche Forschungsgemeinschaft (Germany) through project EJ 7/2-1.

\bibliographystyle{apsrev4-2}
%


\end{document}